\documentclass{article}
\usepackage{spconf,amsmath,graphicx,amssymb, enumitem, setspace}
\usepackage{upgreek}
\usepackage{subfigure}
\usepackage[table,xcdraw]{xcolor}
\usepackage{array,multirow}
\usepackage{multirow,bigdelim}
\usepackage{bm}

\usepackage{amsmath}
\usepackage{hyperref}

\input{def.set}

\usepackage{fancyhdr}
\fancypagestyle{firstpage}{%
\pagestyle{fancy}
  \fancyhf{}
  
\rhead{\footnotesize{}}
\lhead{\footnotesize{
Accepted in Proceedings of the IEEE International Conference on Acoustics, Speech, and Signal Processing (ICASSP)
, Barcelona, Spain, May 4-8, 2020.}}
}

\title{Efficient And Scalable Neural Speech Coding With\\ Collaborative Quantization}
\title{Efficient And Scalable Neural Residual Waveform Coding \\ With Collaborative Quantization}

\name{Kai Zhen$^{1,2}$, Mi Suk Lee$^3$, Jongmo Sung$^3$, Seungkwon Beack$^3$, Minje Kim$^{1,2}$\thanks{This work was supported by Institute for Information \& Communications Technology Promotion (IITP) grant funded by the Korea government (MSIT) (2017-0-00072, Development of Audio/Video Coding and Light Field Media Fundamental Technologies for Ultra Realistic Tera-media).}}
\address{$^1$Indiana University, Luddy School of Informatics, Computing, and Engineering, Bloomington, IN\\
$^2$Indiana University, Cognitive Science Program, Bloomington, IN\\
$^3$Electronics and Telecommunications Research Institute, Daejeon, South Korea\\
  {\small \tt zhenk@iu.edu, lms@etri.re.kr, jmseong@etri.re.kr, skbeack@etri.re.kr, minje@indiana.edu, }}
\begin{document}
\ninept
\maketitle
\thispagestyle{firstpage}

\begin{abstract}
Scalability and efficiency are desired in neural speech codecs, which supports a wide range of bitrates for applications on various devices. We propose a collaborative quantization (CQ) scheme to jointly learn the codebook of LPC coefficients and the corresponding residuals. CQ does not simply shoehorn LPC to a neural network, but bridges the computational capacity of advanced neural network models and traditional, yet  efficient and domain-specific digital signal processing methods in an integrated manner. We demonstrate that CQ achieves much higher quality than its predecessor at 9 kbps with even lower model complexity. We also show that CQ can scale up to 24 kbps where it outperforms AMR-WB and Opus. As a \textcolor{black}{neural} waveform codec, CQ models are with less than 1 million parameters, significantly less than many other generative models.
\end{abstract}
\begin{keywords}
Speech coding, linear predictive coding, deep neural network, residual learning  
\end{keywords}

\section{Introduction}
Speech coding quantizes speech signals into a compact bit stream for efficient transmission and storage in telecommunication systems  \cite{brandenburg1994iso, hasegawa2003speech}. 
The design of speech codecs is to address the trade-off among low bitrate, high perceptual quality, low complexity and delay, etc  \cite{gibson2005speech, choudhary2014study}.
Most speech codecs are classified into two categorizes, \textcolor{black}{\em vocoders}
and {\em waveform} coders \cite{spanias1994speech}. Vocoders use few parameters to model the human speech production process, such as vocal tract, pitch frequency, etc  \cite{ItakuraF1990lpc}. In comparison, waveform coders compress and reconstruct the waveform to make the decoded speech similar to the input as ``perceptually" as possible. Conventional vocoders are computationally efficient and can encode speech at very low bitrates, while waveform coders support a much wider bitrate range with scalable performance and are more robust to noise.

In both conventional vocoders and waveform coders, linear predictive coding (LPC) \textcolor{black}{\cite{o1988linear}}, an all-pole linear filter, serves a critical component, as it can efficiently model power spectrum with only a few coefficients through Levinson-Durbin algorithm  \cite{ItakuraF1990lpc}. For vocoders, the LPC residual is then modeled as a synthetic excitation signal with a pitch pulse train or white noise component  \cite{moriya2016progress}. On the other hand, for waveform coders, such as Opus  \cite{valin2016highopus}, Speex  \cite{speex} and AMR-WB  \cite{ITUamrwb}, the residual is directly compressed to the desired bitrate before being synthesized to the decoded signal.

LPC is useful in modern neural speech codecs, too. While generative autoregressive models, such as WaveNet, have greatly improved the synthesized speech quality  \cite{van2016wavenet}, it comes at the cost of model complexity during the decoding process  \cite{KleijnW2018wavenet}. For example, vector quantized variational autoencoders (VQ-VAE) with WaveNet decoder achieves impressive speech quality at a very low bitrate of 1.6 kbps, 
yet with approximately 20 million trainable parameters  \cite{GarbaceaC2019vqvae}. To make such a system more efficient, LPC can still unload computational overheads from neural networks. LPCNet combines WaveRNN  \cite{kalchbrenner2018efficient} and LPC to shrink down the complexity to 3 GFLOPS which enables real-time coding  \cite{ValinJ2019lpcnet, valin2019real}. \textcolor{black}{Nevertheless, }LPCNet, as a vocoder,  provides a decent performance at 1.6  kbps, but does not scale up to transparent quality. \textcolor{black}{In terms of the neural waveform coder, CMRL \cite{ZhenK2019interspeech} uses LPC as a pre-processor and a variation of \cite{KankanahalliS2018icassp} to model the LPC residual to match the state-of-the-art speech quality with only 0.9 million parameters. However, both LPCNet and CMRL take LPC another blackbox shoehorned into advanced neural networks. Using LPC as a deterministic pre-processor can be sub-optimal, as its bit allocation is pre-defined and not integrated to model training.} 


To better incorporate LPC with neural networks towards scalable waveform coding with low model complexity, we propose a collaborative quantization (CQ) scheme where LPC quantization process is trainable. Coupled with the other neural network autoencoding modules for the LPC residual coding, the proposed quantization scheme learns the optimal bit allocation between the LPC coefficients and the other neural network code layers. With the proposed collaborative training scheme, CQ outperforms its predecessor at 9 kbps, and can scale up to match the performance of the state-of-the art codec at 24 kbps with a much lower complexity than many generative models. We first illustrate relevant techniques which CQ is based upon in Section \ref{sec:relatedwork}, and then explain how they are tailored to our model design in Section \ref{sec:model}. In Section \ref{sec:exp}, we evaluate the model in multiple bitrates in terms of objective \textcolor{black}{and subjective}  measures. We conclude in Section \ref{sec:conclusion}.

\begin{table}[t]
\centering
\footnotesize
\vspace{-.2in}
\caption{Architecture of the 1D-CNN autoencoders. Input and output tensors sizes are represented by (width, channel), while the kernel shape is (width, in channel, out channel).}
\setlength\tabcolsep{5.6pt}
\begin{tabular}{ c|c|c|c }
 \hline
 Layer &Input shape & Kernel shape & Output shape\\
 \hline
Change channel & (512, 1) & (9, 1, 100) &(512, 100) \\
\hline
1st bottleneck & (512, 100) & \begin{tabular}{cc}\rule[6pt]{0pt}{0pt}(9, 100, 20)  &\rdelim]{3}{5mm}[$\times$2]\\ (9, 20, 20)  &  \\(9, 20, 100) &\rule[-1pt]{0pt}{0pt}\end{tabular} &(512, 100)  \\\hline
Downsampling & (512, 100) & (9, 100, 100) &(256, 100) \\
\hline
2nd bottleneck & (256, 100) & \begin{tabular}{cc}\rule[6pt]{0pt}{0pt}(9, 100, 20)  &\rdelim]{3}{5mm}[$\times$2]\\ (9, 20, 20)  &  \\(9, 20, 100) &\rule[-2pt]{0pt}{0pt}\end{tabular} &(256, 100)  \\
\hline
Change channel & (256, 100) & (9, 100, 1) &(256, 1) \\
\hline
\hline
Change channel & (256, 1) & (9, 1, 100) &(256, 100) \\
\hline
1st bottleneck & (256, 100) & \begin{tabular}{cc}\rule[6pt]{0pt}{0pt}(9, 100, 20)  &\rdelim]{3}{5mm}[$\times$2]\\ (9, 20, 20)  &  \\(9, 20, 100) &\rule[-2pt]{0pt}{0pt}\end{tabular} &(256, 100)  \\\hline
Upsampling & (256, 100) & (9, 100, 100) &(512, 50) \\
\hline
2nd bottleneck & (512, 50) & \begin{tabular}{cc}\rule[6pt]{0pt}{0pt}(9, 50, 20)  &\rdelim]{3}{5mm}[$\times$2]\\ (9, 20, 20)  &  \\(9, 20, 50) &\rule[-2pt]{0pt}{0pt}\end{tabular} &(512, 50)  \\\hline
Change channel & (512, 50) & (9, 50, 1) &(512, 1) \\
\hline
\end{tabular}
\vspace{-0.05in}
\label{tab:topo}
\end{table}

\section{Preliminaries}
\label{sec:relatedwork}
\subsection{End-to-end speech coding autoencoders}\label{sec:1dcnn}

A 1D-CNN architecture on the time-domain samples serves the desired lightweight autoencoder (AE) for end-to-end speech coding, where the model complexity is a major concern  \cite{KankanahalliS2018icassp, ZhenK2019interspeech}. As shown in Table \ref{tab:topo}, the encoder part consists of four bottleneck ResNet stages  \cite{HeK2016cvpr}, a downsampling convolutional layer to halve the feature map size in the middle, and then a channel compression layer to create a real-valued code vector of 256 dimensions. The decoder is with a mirrored architecture, but its upsampling layer recovers the original frame size (512 samples) from the reduced code length (256).


\subsection{Soft-to-hard (softmax) quantization}\label{sec:softmax}

\textcolor{black}{To compress speech signals}, a core component of this AE is the trainable quantizer which learns a discrete representation of the code layer in the AE. Out of the recent neural network-compatible quantization schemes, such as VQ-VAE  \cite{OordA2017vqvae} and soft-to-hard quantization  \cite{AgustssonE2017softmax}, we focus on soft-to-hard quantization, namely {\em softmax} quantization as in the other end-to-end speech coding AEs  \cite{KankanahalliS2018icassp, ZhenK2019interspeech}. Given an input frame $\bx\in\Real^S$ of $S$ samples, the output from the encoder is $\bh=\mathcal{F}_\text{Enc}(\bx)$, each is a 16-bit floating-point value. Given $J=32$ centroids represented as a vector $\bb\in\Real^J$, softmax quantization maps each sample in $\bh$ to one of $J$ centroids, such that each quantized sample can be represented by $\log_2 J$ bits (5 bits when $J=32$). 

\begin{figure}[t]
\centering\vspace{-0.1in}
\includegraphics[width=0.8\columnwidth]{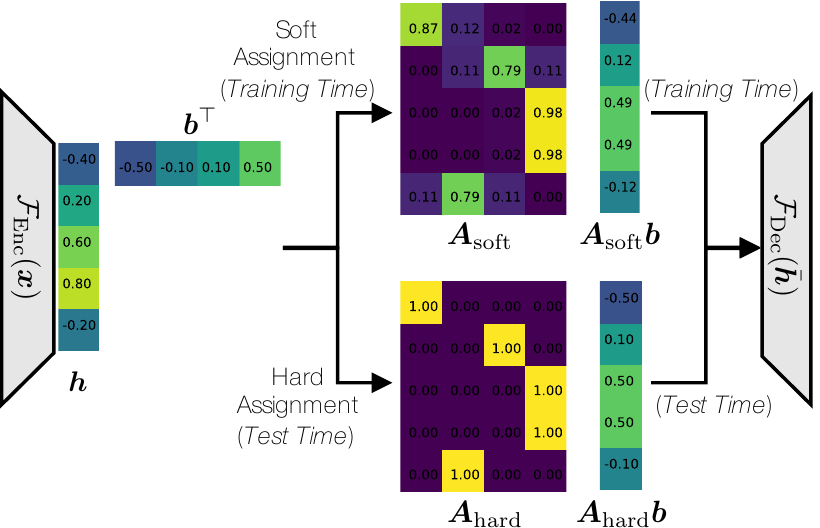}\vspace{-0.1in}
  \caption{An example of the softmax quantization process.} 
  \label{fig:softmax}
\end{figure}

This quantization process uses a hard assignment matrix $\bA_\text{hard}\in\Real^{I\times J}$, where $I$ and $J$ are the dimension of the code and the vector of centroids, respectively. It can be calculated based on the element-wise Euclidean distance matrix $\bD\in\Real^{I\times J}$.
\begin{equation}
\bA_{\text{hard}}(i,j)=\left\{\begin{array}{cl}
1 & \text{ if } \bD(i, j) = \min_{j'} \bD(i, j')\\
0 & \text{otherwise}\end{array}\right..
\end{equation}
Then, the quantization can be done by assigning the closest centroid to each of $\bh$'s elements: $\bar{\bh}=\bA_\text{hard}\bb$. However, this process is not differentiable and blocks the backpropagation error flow during training. Instead, a soft-to-hard assignment is adopted as follows:
\setenumerate[0]{label=(\alph*)} 
\begin{enumerate}[leftmargin=0.2in]
    \item Calculate the distance matrix $\bD\in\Real^{I\times J}$ between the elements of $\bh$  and $\bb$.
    \item Calculate the soft-assignment matrix from the dissimilarity matrix using the softmax function $\bA_\text{soft} = \text{softmax}(-\alpha\bD)$, where the softmax function applies to each row of $\bA_\text{soft}$ to turn it into a probability vector, e.g., $\bA_\text{soft}(i,j)$ holds the highest probability iff $\bh_i$ is most similar to $\bb_j$. Therefore, during the training phase $\bA_\text{soft} \bb$ approximates hard assignments and is fed to the decoder as the input code, while still differentiable. The additional variable $\alpha$ controls the softness of the softmax function, i.e., $\lim_{\alpha\rightarrow\infty}\bA_\text{soft}=\bA_\text{hard}$. We use $\alpha=300$ to minimize the gap between $\bA_\text{soft}$ and $\bA_\text{hard}$.
    \item At testing time, $\bA_\text{hard}$ replaces $\bA_\text{soft}$ by turning the largest probability in a row into one and zeroing the others. $\bA_\text{hard}\bb$ creates the quantized code $\bar{\bh}$.
\end{enumerate}
Fig. \ref{fig:softmax} summarizes the softmax quantization process.



\begin{figure}[t]
\centering
\includegraphics[width=0.75\columnwidth]{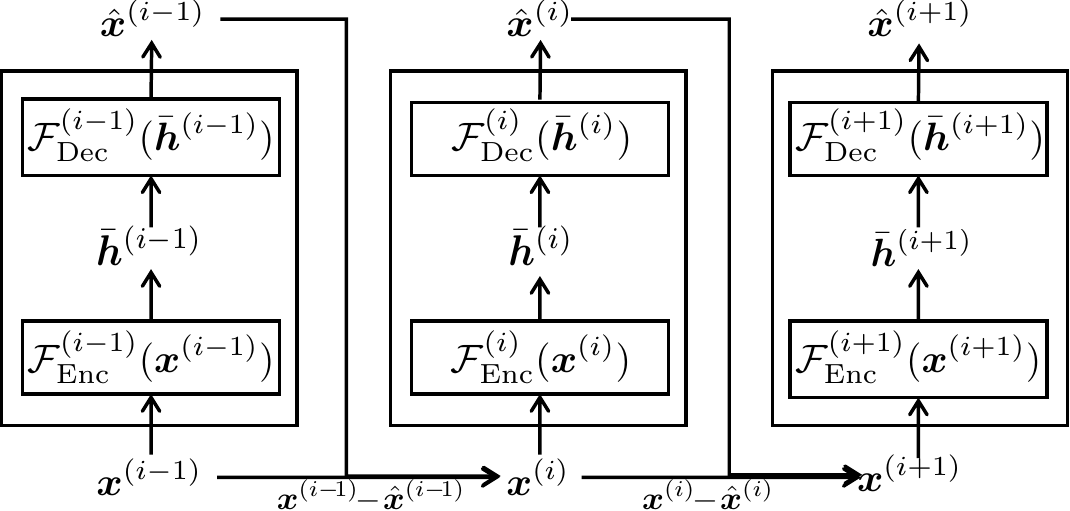}\vspace{-0.1in}
  \caption{The CMRL residual coding scheme.} 
  \label{fig:cmrl}
\end{figure}

\subsection{Cross-module residual learning (CMRL) pipeline} \label{sec:cmrl}
CMRL serializes a list of AEs as its building block modules to enable residual learning among them (Fig. \ref{fig:cmrl}). Instead of relying on one AE,  CMRL serializes a list of AEs as building block modules, where the $i$-th AE takes its own input $\bx^{(i)}$ and is trained to predict it $\hat{\bx}^{(i)}\approx\bx^{(i)}$. Except for the heading AE, the input of $i$-th AE $\bx^{(i)}$ is the residual signal, or the difference between the input speech $\bx$ and the sum of what has not been reconstructed by the preceding AEs: $\bx^{(i)} = \bx - \sum_{j=1}^{i-1}\hat{\bx}^{(j)}$. CMRL decentralizes the effort of optimizing one gigantic neural network; lowers the model complexity in terms of trainable parameters to less than 1 million, which brings neural audio coding algorithms closer to smart devices with limited energy supply and storage space. The AEs in CMRL use the same model architecture in Section \ref{sec:1dcnn} and quantization scheme in Section \ref{sec:softmax}.




\begin{figure*}[t]
\hspace{0in}
\begin{minipage}{.33\textwidth}
\centering
\subfigure[Cross-frame windowing]{\includegraphics[scale=0.55]{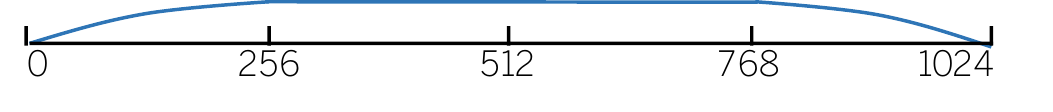}}
\subfigure[Sub-frame windowing]{\includegraphics[scale=0.55]{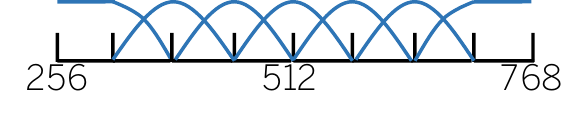}}
\subfigure[Synthesis windowing]{\includegraphics[scale=0.55]{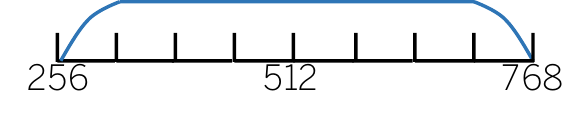}}
\caption{LPC windowing schemes}
\label{fig:windowing}
\end{minipage}
\hspace{-.2in}
\begin{minipage}{.37\textwidth}
\centering
\vspace{0.05in}
\includegraphics[scale=0.46]{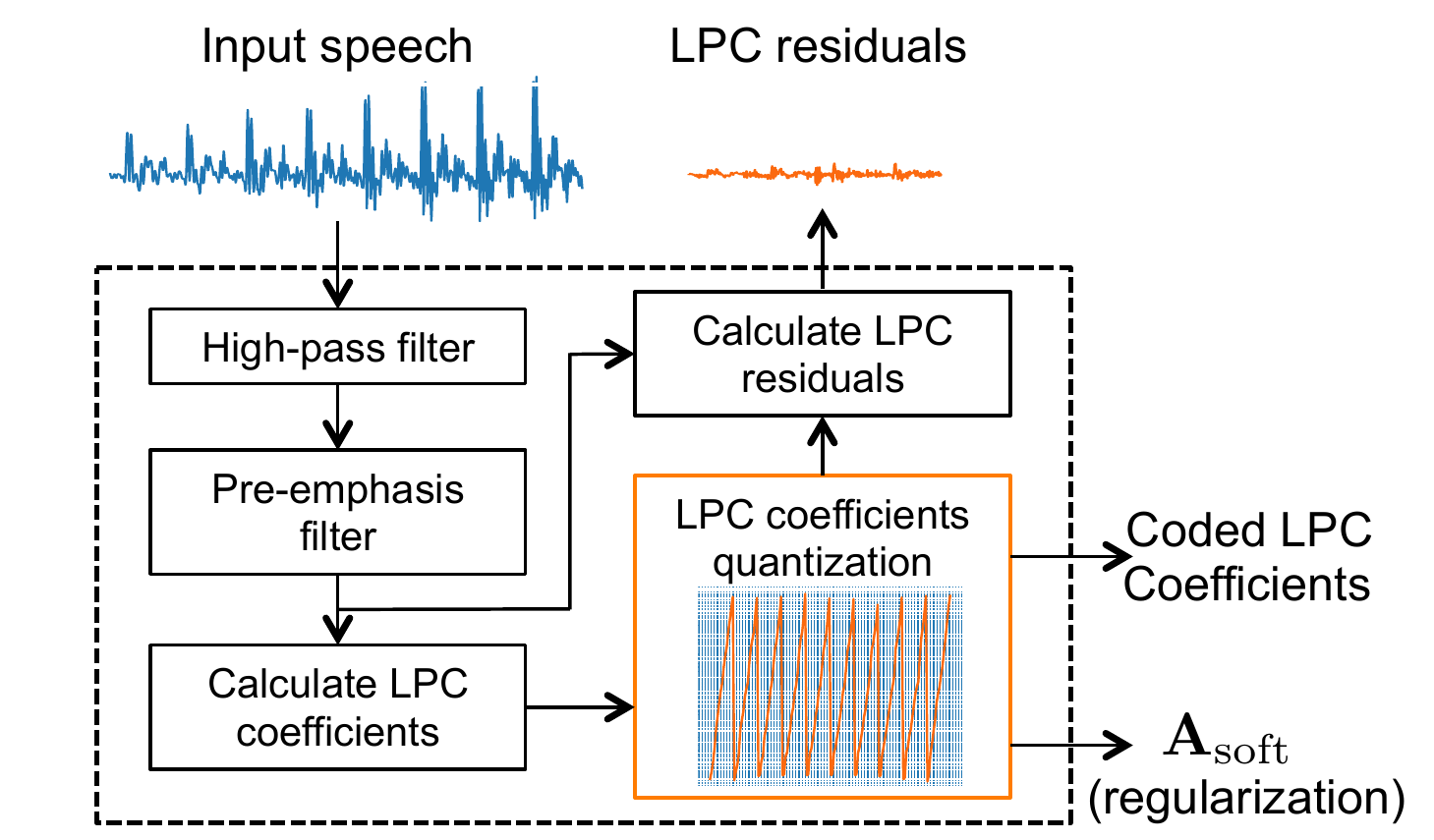}
\caption{The trainable LPC analyzer}
\label{fig:lpc_diagram}
\end{minipage}
\begin{minipage}{.3\textwidth}
\centering
\includegraphics[scale=0.45]{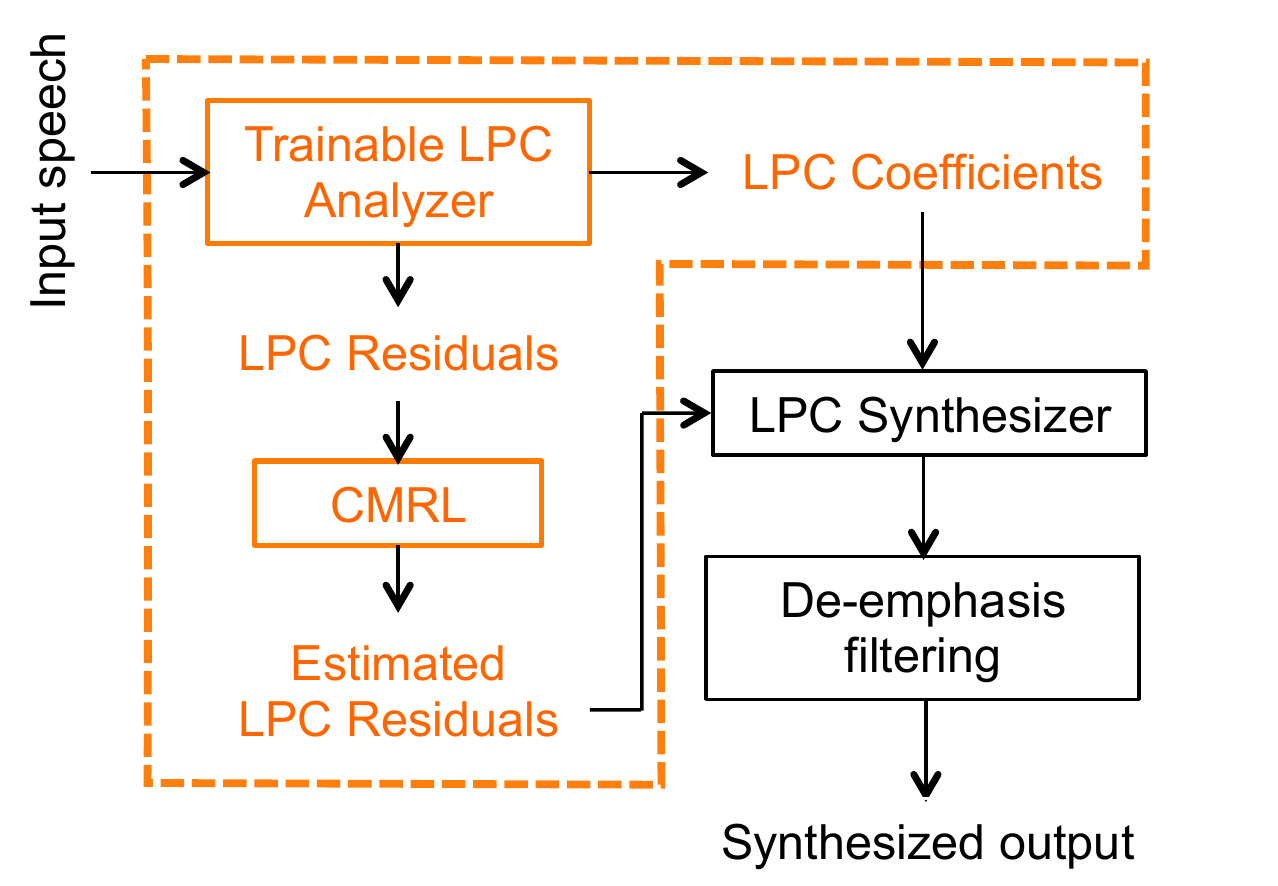}
\caption{Overview of the CQ system.}
\label{fig:cq_diagram}
\end{minipage}
\end{figure*}

\section{Collaborative quantization}
\label{sec:model}


In the CMRL pipeline, LPC module serves a pre-processor with a fixed bitrate of 2.4 kbps. While it can effectively model the spectral envelope, it may not fully benefit the consequent residual quantization. For example, for a frame that LPC cannot effectively model, CQ can weigh more on the following AEs to use more bits, and vice versa. In this section, we break down the LPC process to make its quantization module trainable, along with the other AE modules in CMRL which are to recover the LPC residual as best as possible.  

\subsection{Trainable LPC analyzer}


Our goal is to incorporate LPC analysis into the CMRL pipeline so that it outsources the LPC coefficient quantization to the neural network training algorithm. 
The trainable LPC analyzer is derived from  AMR-WB \cite{BessetteB2002amrwb} with several necessary adjustments to be compatible with neural network computational paradigm.

\noindent{\em \textbf{High-pass filtering and pre-emphasizing}}: Given the input speech, we first adopt high-pass filtering and pre-emphasizing as in  \cite{BessetteB2002amrwb}. A high-pass filter 
is employed with a cut off frequency of 50 Hz. The pre-emphasis filter is $H_{emp}(z) = 1 - 0.68z^{-1}$, and the de-emphasis filter is employed to remove artifacts in the high frequencies. 

\noindent{\em \textbf{Data windowing for LPC coefficients calculation}}: The pre-emphasis filtered utterances are segmented to frames of 1024 samples. Each frame is windowed before LPC coefficients are calculated. As shown in Fig. \ref{fig:windowing} (a), the symmetric window has its weight emphasized on the middle 50\% samples: first 25\% part is the left half of a Hann window with 512 points; the middle 50\% is a series of ones; and the rest 25\% part is the right half of the Hann window. Then, the linear prediction is conducted on the windowed frame in time domain  $s$. For the prediction of the $t$-th sample, 
$\hat{s}{(t)}=\sum_{i}a_{i}s{(t-i)}$,
where $a_i$ is the $i$-th LPC coefficient. 
The frames are with 50\% overlap. The LPC order is set to be 16. We use Levinson Durbin algorithm  \cite{ItakuraF1990lpc} to calculate LPC coefficients. They are are represented as line spectral pairs (LSP)  \cite{soong1984line} which are more robust to quantization.


\noindent{\em \textbf{Trainable LPC quantization}}: We then employ the trainable softmax quantization scheme to LPC coefficients in LSP domain, to represent each coefficient with its closest centroid, as described in Section \ref{sec:softmax}. For each windowed frame $\bx$, $\bh_\text{LPC}=\mathcal{F}_{LPC}(\mathbf{x})$ gives corresponding LPC coefficients in the LSP representation. The rest of the process is the same with the softmax quantization process, although this time the LPC-specific centroids $\bb_\text{LPC}$ should be learned and be used to construct the soft assignment matrix. In practice, we set LPC order to be 16, and the number of centroids to be 256 (i.e., 8 bits). Hence, the size of the soft and hard assignment matrices is 16$\times$256, each of whose rows is a probability vector and a one-hot vector, respectively.

\noindent{\em \textbf{Data windowing for LPC residual calculation}}: We use a sub-frame windowing technique to calculate residuals (Fig. \ref{fig:windowing} (b)). For a given speech frame and its quantized LPC coefficients, we calculate residuals for each sub-frame, individually. The middle 50\% of the 1024 samples, for example, [256:768] for the first analysis frame that covers [0:1024] and [768:1280] for the second frame of [512:1536], is decomposed into seven sub-frames, each with the size 128 and 50\% overlap. Out of the seven sub-frames, the middle five are windowed by a Hann function with 128 points; the first and last frames are asymmetrically windowed, as shown in Fig. \ref{fig:windowing} (b). The residual is calculated with the seven sub-frames on the middle 512 samples, which amount to 50\% of the frame. Hence, given the 50\% analysis frame overlap, there is no overlap between residual segments.

%


\begin{figure}[t]
\centering\vspace{-0.1in}
\includegraphics[width=0.85\columnwidth]{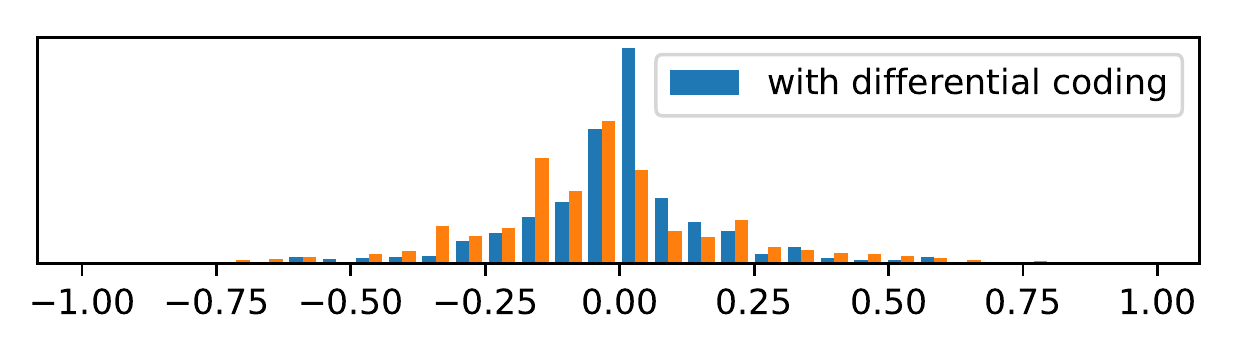}\vspace{-0.2in}
  \caption{Differential coding enables a more centralized distribution} 
  \label{fig:dc}
\end{figure}

\subsection{Residual coding}
The LPC residual, calculated from the trainable LPC analyzer (Fig. \ref{fig:lpc_diagram}), is compressed by the 1D-CNN AEs as described in Section \ref{sec:cmrl}. In this work, we employ differential coding  \cite{cummiskey1973adaptive} to the output of encoders, $\bh=[h_0, h_1, \cdots, h_{m-1}]$ where $m$ is the length of code per frame for each AE. Hence, the input scalar to the softmax quantization is $\Updelta h_i=h_i-h_{i-1}$. Consequently, the quantization starts from a more centralized real-valued ``code" distribution (Fig.\ref{fig:dc}). 
As illustrated in Fig. \ref{fig:cq_diagram}, both the quantization of LPC coefficients and residual coding with CMRL are optimized together. With this design, the purpose of LPC analysis is not just to minimize the residual signal energy as much as possible  \cite{paliwal2012vector}, but to find a pivot which also facilitates the residual compression from following CMRL modules. 



\subsection{Model training}

According to the CMRL pipeline, the individual AEs can be trained sequentially by using the residual of the previous module as the input of the AE and the target of prediction. Once all the AEs are trained, finetuning step follows to improve the total reconstruction quality. This section discusses the loss function we used for training each of the AEs as well as for finetuning. The loss function consists of the reconstruction error terms and regularizers:
\begin{equation}\label{eq:object}
\mathcal{L} = \lambda_1\mathcal{T}(y||\hat{y}) +  \lambda_2\mathcal{F}(y||\hat{y}) + \lambda_3\mathcal{Q}(\bA_\text{soft}) + \lambda_4\mathcal{E}(\bA_\text{soft})
\end{equation}

Given that the input of CQ is in time domain, we minimize the loss in both time and frequency domains. The time domain error, $\mathcal{T}(y||\hat{y})$, is measured by mean squared error (MSE). $\mathcal{F}(y||\hat{y})$ compensates what is not captured by the non-perceptual $\mathcal{T}(y||\hat{y})$ term by measuring the loss in mel-scale frequency domain. Four different mel-filter banks are specified with the size of 128, 32, 16 and 8, to enable a coarse-to-fine differentiation.

$\mathcal{Q}(\bA_\text{soft})$ and $\mathcal{E}(\bA_\text{soft})$ are regularizers for softmax quantization. For the soft assignment matrix $\bA_\text{soft}$ as defined in Section \ref{sec:softmax}, $\mathcal{Q}(\bA_\text{soft})$ is defined as $\sum_{i,j}(\sqrt{\bA_\text{soft}(i,j)}-1)/I$, to ensure that the soft assignment is close to its hard assignment version. 
$\calE(\bA_\text{soft})$ calculates the entropy of softmax-quantized bit strings, to control the bitrate. First, the frequency of each kernel is calculated by summing the rows of the soft assignment matrix:  $\bA_\text{soft} (\cdot, j) =\sum_{i}\bA{(i,j)}$. The corresponding probability distribution over the kernels, denoted as $\bp$, is on how often a code is assigned to each kernel: $p_j = \bA(\cdot, j)/ (IJ)$. The entropy is, therefore,  $\mathcal{E}(\bA_\text{soft}) =-\sum_j p_j\log_{2}(p_j)$. As firstly proposed in  \cite{KankanahalliS2018icassp}, by adjusting $\lambda_4$, the model is finetuned to the range of the desired bitrate. We find that applying Huffman coding on grouped sample pairs (two adjacent samples per pair) gives a better compression ratio, as it further utilizes the temporal structure preserved in the quantized residual signals. Note that there is no need to allocate bits for model parameters as in autoregressive models \cite{kleijn2007rate}.

\section{Experiments}
\label{sec:exp}
\subsection{Experimental Settings}
We consider four bitrate cases 9, 16, 20, and 24  kbps, with the sample rate of 16 kHz. The training set contains 2.6 hours of speech from 300 speakers randomly selected from the TIMIT training set. 50 speakers are randomly selected from the test set. At test time, each frame has 512 samples with an overlap of 32 samples. The overlap region is windowed by Hann function (Fig.\ref{fig:windowing}(c)).
For 24 kbps, we cascade the LPC module and two AEs as in \cite{ZhenK2019interspeech}, but we use only one AE for the LPC residual coding for other three bitrates. For 16 and 20  kbps cases, the code layer is downsampled with a convolutional layer of stride 2; for the 9 kbps case, we use two downsampling layers of stride 2. We use Adam optimizer \cite{adam} with the batch size of 128, learning rate of 0.0002 for 30 epochs, followed by finetuning until the entropy is within the target range.



\begin{table}[]
\centering
\caption{MOS-LQO scores computed from PESQ-WB}
\footnotesize
\begin{tabular}{c|c|c|c|c}
\hline
    & AMR-WB & Opus & LPC-CMRL & CQ   \\ \hline
\multicolumn{1}{l|}{$\sim$9 kbps}  & 3.48   & 3.42 & 3.01     & 3.69 \\ \hline
\multicolumn{1}{l|}{$\sim$16 kbps}  & 3.99   & 4.30 & 3.26     & 3.98 \\ \hline
\multicolumn{1}{l|}{$\sim$20 kbps}  & 4.09   & 4.43 & 3.67     & 4.08 \\ \hline
\multicolumn{1}{l|}{$\sim$24 kbps} & 4.17   & 4.47 & 4.15     & 4.17 \\ \hline
\end{tabular}
\label{tab:pesq}
\end{table}


Recent works on generative model based speech synthesis systems \cite{ValinJ2019lpcnet,KleijnW2018wavenet} have reported that PESQ \cite{rix2001perceptual} or its successor POLQA \cite{beerends2013perceptual} cannot accurately evaluate the synthesized speech quality. In fact, we also find that there is a discrepancy between PESQ and the actual MOS. 
Still, we report MOS-LQO scores in Table \ref{tab:pesq} as the proposed method is based on waveform reconstruction.

We conduct two MUSHRA-like \cite{mushra} sessions corresponding to two bitrate settings. Each session includes ten trials on gender-balanced utterances randomly chosen from the test set. A low-pass anchor at 4kHz and the hidden reference signal are included in both settings, with eleven audio experts as the subjects. The lower bitrate setting refers to the performance around 9 kbps, including AMR-WB \cite{BessetteB2002amrwb} at 8.85  kbps, Opus \cite{valin2016highopus} at 9 kbps, LPC-CMRL \cite{ZhenK2019interspeech} at 9 and 16  kbps, and CQ at 9 kbps. The higher bitrate session uses decoded signals from codes with around 24 kbps bitrate. The competing models are AMR-WB at 23.85  kbps, Opus at 24  kbps, the proposed CQ method, and the LPC-CMRL counterpart at 24 kbps.

\subsection{Experimental Results}

\begin{figure}[t]
\centering
\subfigure[$\sim$9 kbps]{\includegraphics[width=.2\textwidth]{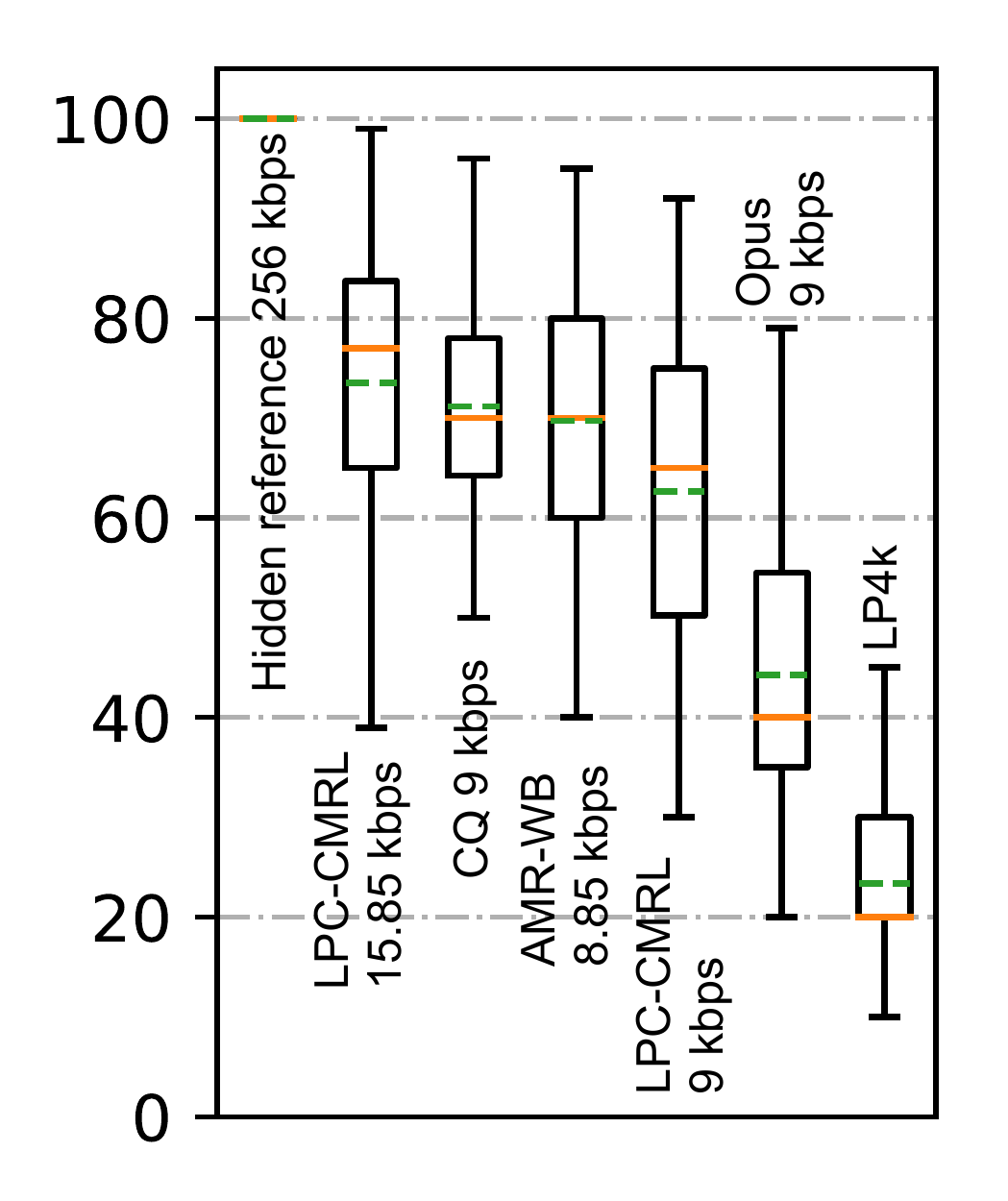}}
\hspace{0.1in}
\subfigure[$\sim$24 kbps]{\includegraphics[width=.2\textwidth]{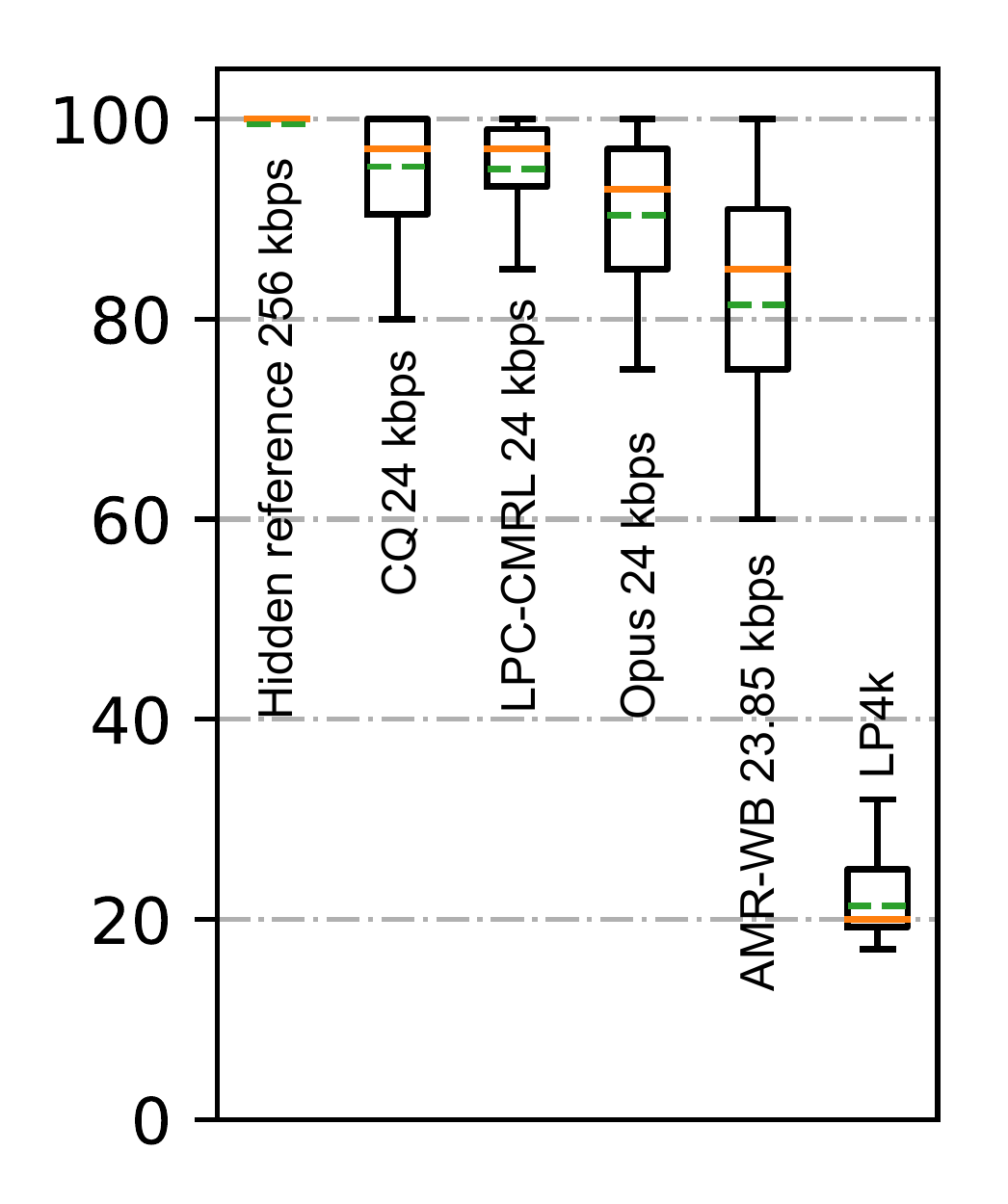}}
  \caption{MUSHRA results in box-plots (Orange solid lines represent medians, and green dashed lines represent means).} 
  \label{fig:resnet}
\end{figure}


First, we can see that CQ outperforms LPC-CMRL at the same bitrate, especially in low bitrate setting (Fig.~\ref{fig:resnet}). 
In higher bitrate setting, both LPC-CMRL and CQ outperform AMR-WB and Opus in the MUSHRA test. None of these methods add very audible artifacts. One explanation for the result is that AMR-WB does not code all 8kHz wide bandwidth, but up to 7kHz, while our model maintains the energy of decoded signals at high frequencies, and therefore yield less-muffled speech. However, as the bitrate decreases, some human subjects tend to be more negative towards the artifacts (from LPC-CMRL) which become audible than the moderately muffled speech (from AMR-WB). That explains why LPC-CMRL is less favored than AMR-WB at low bitrate. As was expected, when the LPC coefficient quantization is collaboratively learned along with residual coding, the artifact is suppressed---CQ-9 outperforms LPC-CMRL-9 with a noticeable margin\footnote{Decoded samples are available at 
\url{https://saige.sice.indiana.edu/research-projects/neural-audio-coding/}}.



\subsection{Complexity}
We use the number of trainable parameters to measure the model complexity, as it determines the floating point operation rate, memory space, and execution time, etc. Each AE, used in CQ and CMRL, contains 0.45 million parameters (Table \ref{tab:topo}). We use one AE for CQ-16 and CQ-20 kbps cases, and two AEs for the 24 kbps case. The AE in CQ-9 kbps is 0.67 million parameters as it contains two downsampling and upsampling layers. Admittedly, the decoder of CQ is still more complex than the conventional codecs, but it is much simpler than the other WaveNet vocoder-based coding systems 
, e.g., the VQ-VAE with WaveNet has 20 million parameters, although it gives impressive speech quality with only 1.6 kbps. 



\section{Conclusion}
\label{sec:conclusion}
We proposed a lightweight and scalable waveform neural codec, integrading merits from both advanced end-to-end neural network architectures and conventional DSP techniques. With collaborative quantization (CQ), LPC coefficient quantization becomes a trainable component to be jointly optimized with the residual quantization. 
This helps CQ outperform its predecessor and Opus at ~9 kbps, and show comparable performance to AMR-WB at 8.85 kbps. The method is with much lower complexity in terms of the amount of parameters than other competitive neural speech coding models. The source code is publicized under a MIT
license\footnote{Available at 
\url{https://github.com/cocosci/NSC/}}.

\bibliographystyle{IEEEbib}
\bibliography{strings,refs19,mjkim}

\end{document}